\documentstyle[psfig]{mn0}

\begin{document}

\title[Optical afterglows]
 {Optical afterglows of Gamma-Ray Bursts: peaks, plateaus, and possibilities}

\author[Panaitescu \& Vestrand]{A. Panaitescu and W.T. Vestrand \\
       Space Science and Applications, MS D466, Los Alamos National Laboratory,
       Los Alamos, NM 87545, USA}

\maketitle

\begin{abstract}
\begin{small}
 \\
 The optical light-curves of GRB afterglows display either peaks or plateaus. 
 We identify 16 afterglows of the former type, 17 of the latter, and 4 with broad
peaks, that could be of either type. The optical energy release of these two classes
is similar and is correlated with the GRB output, the correlation being stronger 
for peaky afterglows, which suggests that the burst and afterglow emissions of 
peaky afterglows are from the same relativistic ejecta and that the optical emission 
of afterglows with plateaus arises more often from ejecta that did not produce the 
burst emission. 
 Consequently, we propose that peaky optical afterglows are from impulsive ejecta 
releases and that plateau optical afterglows originate from long-lived engines,
the break in the optical light-curve (peak or plateau end) marking the onset of the 
entire outflow deceleration. 
 In the peak luminosity--peak time plane, the distribution of peaky afterglows displays 
an edge with $L_p \propto t_p^{-3}$, which we attribute to variations (among afterglows) 
in the ambient medium density. The fluxes and epochs of optical plateau breaks follow 
a $L_b \propto t_b^{-1}$ anticorrelation.
 Sixty percent of 25 afterglows that were well-monitored in the optical and X-rays 
show light-curves with comparable power-law decays indices and achromatic breaks. 
The other 40 percent display three types of decoupled behaviours: 
$i)$ chromatic optical light-curve breaks (perhaps due to the peak of the synchrotron 
     spectrum crossing the optical), 
$ii)$ X-ray flux decays faster than in the optical (suggesting that the X-ray emission 
    is from local inverse-Compton scattering), and 
$iii)$ chromatic X-ray light-curve breaks (indicating that the X-ray emission is from 
   external up-scattering).
\end{small}
\\ 
\end{abstract}


\section{Introduction}

 The {\sl prompt} emission of Gamma-Ray Bursts (GRBs) is thought to be produced by particles 
energized in internal shocks occurring in an unsteady outflow (Rees \& M\'esz\'aros 1994, 
Ramirez-Ruiz \& Fenimore 2000). The interaction of the GRB outflow with the circumburst medium
generates shocks that accelerate energetic particles which, in turn, radiate the so-called 
{\sl afterglow} emission (M\'esz\'aros \& Rees 1997). The spectral, temporal, and polarization 
properties of that afterglow are fundamental observational probes of the outflow interaction 
with the surrounding medium.

 The X-ray afterglows observed by Swift display light-curves with up to two inflection points 
or "breaks" (e.g. Nousek et al 2006, O'Brien et al 2006, Zhang et al 2006). A minority of afterglow 
light-curves exhibit a single power-law decay for decades in time, while the majority display 
an initial phase of slow flux decay (a "plateau") lasting up to 1-10 ks after trigger, followed 
by a more rapid decline. While generally consistent with the expectations for synchrotron emission 
from an adiabatic forward-shock energizing the circumburst medium, 
there are cases where the decay is slower than expected for that model (Willingale et al 2007).
Those slower-than-expected decays can be due to an interval of energy injection that powers
the blast-wave. Some of the X-ray afterglows monitored for very long durations display a second 
light-curve break that could be a jet-break (e.g. Panaitescu 2007, Racusin et al 2010), although 
some of those breaks do not satisfy the expected closure relation between the flux decay index 
and spectrum slope (Liang et al 2008).

 At gamma-ray energies, the prompt emission dominates the observed flux. In the X-rays, the 
fast-decaying tail of the prompt emission is also dominant up to hundreds of seconds after the 
trigger (e.g. Tagliaferri et al 2005). This prominence of the prompt X-ray emission makes it difficult 
to detect the X-ray afterglow light-curve peak produced either when the blast-wave begins to 
decelerate (i.e. when the reverse-shock has crossed all the GRB ejecta) or when the peak of 
the synchrotron spectrum traverses the X-ray band. Prompt emission has also been detected at 
optical wavelengths (e.g. Akerlof et al 1999, Vestrand et al 2005, Racusin et al 2008, Wozniak 
et al 2009).  However, in the optical, the prompt emission is usually less prominent, making it 
possible sometimes to detect the peak of the optical afterglow (e.g. Molinari et al 2007) even 
when the prompt component is present (e.g. Vestrand et al 2006).  

 Panaitescu \& Vestrand (2008) have studied the morphology of optical afterglow light-curves
and have identified two prominent classes, of afterglows showing early peaks and afterglows 
showing extended plateaus. We have suggested that these two behaviors resulted from the observer 
being located initially outside the jet aperture (for peaked afterglows) and from outflows having 
a non-uniform angular distribution of the ejecta kinetic energy per solid angle (for afterglows 
with plateaus).

 In this work, we expand our earlier study by adding new optical light-curves that display peaks 
and plateaus, and continue to investigate the possible reasons for the observed diversity of early 
optical light-curves (section 2). In particular, we explore another potential explanation: 
the onset of the blast-wave deceleration. In section 3, we examine the correlation of optical 
and X-ray light-curve and discuss the mechanisms that can account for the diverse optical and X-ray 
light-curve relative behaviours.

\section{Possible origins for light-curve peaks and plateaus}

\begin{figure}
\centerline{\psfig{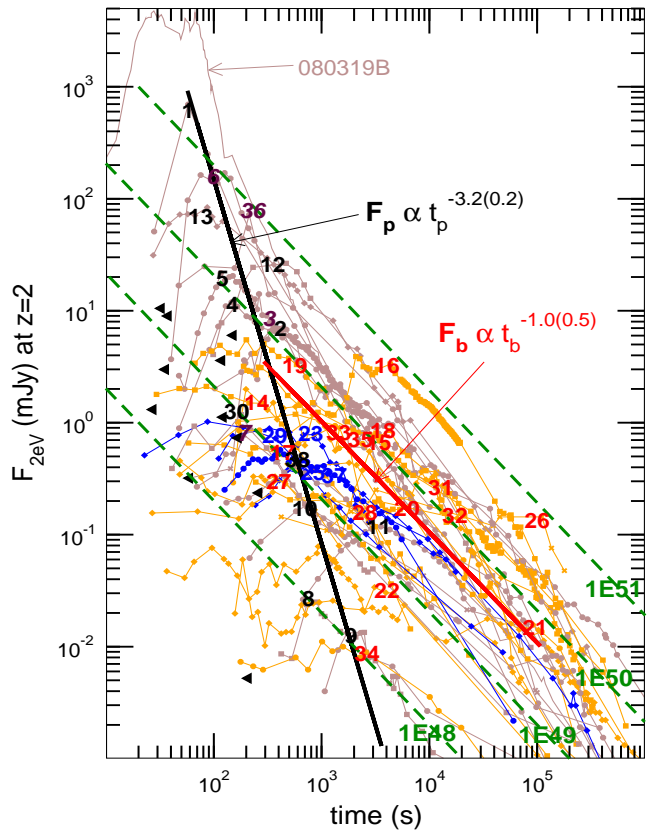}}
\caption{ Light-curves of 16 afterglows with optical peaks (brown curves, peak location shown 
   with black numbers), 17 afterglows with optical plateaus (orange curves, peak location shown 
   with red numbers), and 4 afterglows of uncertain type, with either a broader peak or a shorter
   plateau (blue lines and numbers). These GRB afterglows are listed in Table 1.
   All light-curves are for afterglows moved at redshift $z=2$.
   Numbers 3, 6, 7, and 36 indicate the peaks of afterglows 050820, 061007, 061121, and 050904,
   respectively, which occurred simultaneously with a burst pulse, thus their peaks may have 
   been produced by the prompt emission mechanism and not by the afterglow (as for GRB 080319B, 
   whose counterpart optical emission tracks, although not perfectly, the burst fluctuations). 
   Power-law fits to peaks and plateau ends are shown with black and red lines, respectively.
   Green dashed lines indicate a fixed energy release (in ergs). 
   Black triangles indicate the first measurement of optical afterglows that display a decaying 
   light-curve (no peak or plateau); thus the triangles represent an upper limit on the missed 
   peak/plateau end epoch and a lower limit on the peak/plateau flux.  }
\label{z2}
\end{figure}

 Figure \ref{z2} shows the optical light-curves, transformed to a fiducial redshift $z=2$, for 
all afterglows with {\sl $i)$ known redshift} and {\sl $ii)$ sufficiently good optical coverage }
to allow a reliable identification of a light-curve peak or a plateau.
A "peak" is defined by a full-width at half maximum less than a factor of 5 in time. 
A "plateau" is defined by the optical flux displaying a systematic change by a factor less than 
3 over more than a decade in time, ending with a break and followed by a steeper decline. 
The above two criteria are the only used in selecting the sample listed in Table 1.
With these selection criteria, we found 16 afterglows with optical light-curve peaks and 17 with
plateaus. Four afterglows display a broader peak or a shorter-lived plateau than defined above; 
their peaks/plateaus are located in the $F-t$ plane around the intersection of the fits to the 
two categories of afterglows. 

 Peaky afterglows display a strong anticorrelation of the peak flux $F_p$ with the peak epoch 
$t_p$ (linear correlation coefficient $r(\log F_p, \log t_p) = -0.83$,
corresponding to a probability of obtaining a stronger correlation of $10^{-4.4}$ in the
null hypothesis). The afterglows with plateaus exhibit a weaker anticorrelation of the plateau 
flux $F_b$ with the plateau end  epoch $t_b$: $r(\log F_b, \log t_b) = -0.49$ 
(chance probability of 2.3 percent). The strength of these correlations is reduced if the optical 
luminosity is scaled to the GRB output or the peak/plateau end epoch is scaled to the burst 
duration. 
 
 The power-law fits shown in Figure \ref{z2} were obtained by minimizing 
$\chi^2 = \sum_i (y_i-ax_i-c)^2/(a^2\sigma^2_{x,i} +\sigma^2_{y,i})$ 
between the linear fit $y = ax+c$ and the measurement sets $(x_i,y_i)$, with the
 variables $x_i$ and $y_i$ being $\log t_{p,b}$ and $\log F_{p,b}$, respectively, and $\sigma$ being the
 uncertainty of each variable, in log space. The uncertainties arise mostly from our determination of
 the peak/plateau end locations and much less from the errors of reported afterglow photometry. 
 The slopes shown in Figure \ref{z2} are obtained assuming that the relative errors $\epsilon_F = 
 \sigma_F/F_{p,b}$ and $\epsilon_t =\sigma_t/t_{p,b}$ in measuring the peak/plateau end 
 flux and epoch are the same for both quantities ($\epsilon_F = \epsilon_t$) and for all afterglows
 (the slopes of the fits are independent of the assumed relative errors because the fits are done 
 in log-log space); the uncertainties (given in parentheses) of those slopes are estimates obtained 
 by varying $\epsilon_F$ and $\epsilon_t$ within plausible ranges. 

\begin{table}
\caption{ Properties of the optical afterglows used in this work (their $z=2$ light-curves are 
   shown in Figure \ref{z2}). }
\begin{tabular}{lcccccccc}
  \hline \hline
   GRB   &  z   & $t_{p,b}$ & $F_{p,b}$ & $\alpha_o$ & $E_\gamma$     \\
         &      &   (ks)    &   (mJy)   &           &($10^{52}$ erg) \\
         &  (1) &    (2)    &    (3)    &   (4)     &       (5)      \\
  \hline \hline
  \multicolumn{6}{l}{\sc Afterglows with peaks} \\
  990123 & 1.60 &  0.050    & 1200      & 1.80(.11) &  100  \\
  050730 & 3.97 &  0.60     &    1.3    & 0.63(.05) &  12  \\
  050820 & 2.61 &  0.47     &    4.2    & 0.91(.01) &  46  \\
  050904 & 6.29 &  0.39     &    3.9    & 1.15(.03) &  38  \\
  060418 & 1.49 &  0.12     &   43      & 1.13(.02) &  8.8 \\
  060607 & 3.08 &  0.18     &   17      & 1.20(.03) &  8.8 \\
  061007 & 1.26 & 0.07-0.11 &  530      & 1.70(.02) &  4.7 \\
  061121 & 1.31 &  0.075    &  10       & 0.82(.02) &  9.3 \\
  070318 & 0.84 &  0.38     &    2.7    & 0.96(.03) &  2.0  \\
  070419 & 0.97 &  0.55     &    0.13   & 0.99(.07) &  0.33 \\
  070802 & 2.45 &  2.20     &    0.010  & 0.84(.15) &  0.72 \\
  071010A& 0.98 &  0.45     &    1.0    & 0.74(.03) &  0.078 \\
  080710 & 0.85 &  2.30     &    0.89   & 1.52(.02) &  0.58  \\
  080810 & 3.35 &  0.11     &   23      & 1.23(.01) &  14  \\ 
  081203 & 2.05 &  0.35     &   33      & 1.50(.02) &  12  \\
  090418 & 1.61 &  0.16     &   1.9     & 1.21(.04) &  6.1 \\
  \hline
  \multicolumn{6}{l}{\sc Afterglows with plateaus} \\
  050801 & 1.56 &  0.25     &    2.7    & 0.13(.04) &  0.39 \\
  060124 & 2.30 &  3.10     &    0.60   & 0.13(.12) &  18   \\
  060206 & 4.05 &  7.80     &    0.65   & var       &  3.4  \\
  060210 & 3.91 &  0.77     &    0.11   & 0.13(.19) &  18   \\
  060526 & 3.22 &  5.50     &    0.29   & 0.04(.04) &  4.4  \\
  060605 & 3.78 &  0.90     &    0.77   & 0.03(.07) &  18   \\     
  060714 & 2.71 &  7.80     &    0.090  & 0.17(.09) &  9.8  \\ 
  060729 & 0.54 &  52       &    0.15   & 0.07(.01) &  0.40 \\
  060904B& 0.70 &  2.9      &    0.31   & var       &  0.45 \\
  071025 & 5.2  &  2.3      &    0.12   & 0.20(.05) &  49   \\
  080129 & 4.35 &  190      &    0.024  & 0.02(.04) &  4.4  \\
  080310 & 2.42 &  2.6      &    0.42   & 0.14(.09) &  4.6  \\
  080319C& 1.95 &  0.50     &    0.30   & var       &  0.99 \\
  080330 & 1.51 &  2.0      &    0.30   & -0.20(.06)&  0.25 \\
  080928 & 1.69 &  15       &    0.30   & var       &  2.9  \\
  090423 & 8.26 &  55       &    0.011  & 0.33(.15) &  4.8  \\
  090510 & 0.90 &  1.8      &    0.040  & -0.2(.1)  &  0.31 \\
  \hline
  \multicolumn{6}{l}{\sc Afterglows of uncertain type}   \\
  070411 & 2.95 &  1.1      &    0.15   &           &  8.3  \\
  071031 & 2.69 &  2.1      &    0.14   &           &  2.1  \\
  080804 & 2.20 &  0.43     &    0.51   &           &  8.4  \\
  090726 & 2.71 &  2.0      &    0.16   &           &  2.1 \\
\hline \hline \\
\end{tabular}
\begin{minipage}{85mm}
{\bf (1)}: burst redshift;
{\bf (2)}: observer-frame epoch of the optical light-curve peak ($t_p$) or plateau end ($t_b$), 
           typical uncertainty 10-20 percent;
{\bf (3)}: observer-frame optical flux at the epoch of the light-curve peak ($F_p$) or plateau end ($F_b$), 
           typical uncertainty is less 10 percent; for afterglows of uncertain type, this is the time
           when a steep power-law flux decay begins; 
{\bf (4)}: exponent of the power-law optical flux decay ($F_o \propto t^{-\alpha_o}$) after the
           peak or during the plateau ($1\sigma$ uncertainty in parentheses), "var" indicates
           substantial variability during the plateau;
{\bf (5)}: GRB output in 10-1000 keV (rest-frame) calculated from fluences, peak energies, 
           and spectral slopes reported in GCNs (typical uncertainty 20 percent). 
           When peak energy ($E_p$) or the low- and high-energy
           spectral slopes are not known, we assumed the most-likely values found by Preece et al (2000)
           for 156 bright BATSE bursts: $E_p = 250$ keV, $\beta_{low} = 0$, $\beta_{high} = 1.4$;
           however, if $E_p$ was not measured and the 15-150 keV GRB spectrum measured by Swift/BAT
           is soft, we assumed $E_p = 25$ keV.
\end{minipage}
\end{table}

 We note that these anticorrelations (in particular that for peaky afterglows) are due, in part, 
to an observational limitation, as dimmer afterglows peaking at earlier times are more likely 
to be missed. Figure \ref{z2} shows mostly afterglows with a peak near the bright edge of the 
$F_p-t_p$ distribution, and is quite likely that future more rapid and deeper follow-ups of GRB
afterglows will fill the lower-left part of Figure \ref{z2} (e.g. afterglow 080430 -- Klotz et al 2009).
Thus, the correlations identified in Figure \ref{z2} define, more precisely, {\sl the boundary 
of a "zone of avoidance"} in upper-right region of that figure.

 The important question is what generates these boundaries ?
 The $F_b \propto t_b^{-1}$ anticorrelation for plateau breaks may only mean that, given the 
energetics of relativistic outflows produced by GRB progenitors, afterglows with an optical 
output larger than about $10^{50}$ erg/sr are not produced. However, the much steeper 
$F_p \propto t_p^{-3}$ dependence displayed by the edge of the $F_p-t_p$ distribution appears 
more likely to arise from the mechanism which produces light-curve peaks, thus {\sl that 
correlation should be used as a tool for investigating the possible light-curve peak mechanisms}. 

\begin{figure*}
\centerline{\psfig{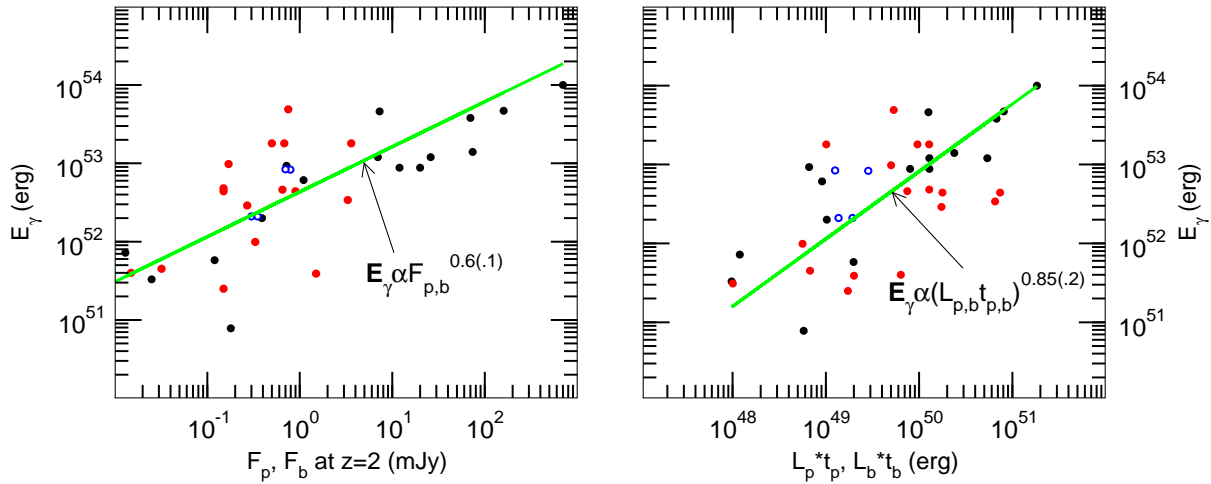}}
\caption{ The GRB output $E_\gamma$ (Table 1) for the afterglows shown in Figure \ref{z2} is 
   correlated to the optical luminosity at the peak/plateau end time (left panel) and with 
   the optical energy released (right panel). 
   Black symbols are for afterglows with optical peaks, red symbols for optical plateaus,
   open circles for afterglows of uncertain type.
   The linear correlation coefficients are $r(\log F_{p,b},\log E_\gamma) = 0.75$ and 
   $r(\log E_o,\log E_\gamma) = 0.66$ for all 37 afterglows, where $E_o = L_{p,b}t_{p,b}$
   and $L_{p,b}$ is the afterglow optical luminosity at $t_p$ or $t_b$. 
   These linear correlation coefficients correspond to a probability for a chance correlation 
   of $10^{-7.3}$ and $10^{-5.3}$, respectively.
   The optical luminosity (or output) of peaky afterglows is much better correlated with the 
   GRB energy release than for plateau afterglows; for instance, $r(\log F_p, \log E_\gamma) 
   = 0.87$ for the former and $r(\log F_b,\log E_\gamma) = 0.57$ for the latter.  }
\label{LpEg}
\end{figure*}

 Figure \ref{LpEg} shows that the GRB output is well correlated with the afterglow optical luminosity
at the peak or plateau break, as well as with the afterglow energy output in the optical. 
{\sl The optical--GRB output correlation is much stronger for peaky afterglows than for plateau 
afterglows}. This result should be tested in the future with a larger set of afterglows, as 
it may provide support to the "deceleration-onset" hypothesis for the origin of afterglow 
light-curve peaks and plateaus. We note that Kann et al (2010) have found a tentative correlation
between the optical luminosity at 1 day (i.e. at a fixed time) and the GRB energy release for a
larger sample of 76 afterglows. Liang et al (2010) have also found a strong correlation of the 
peak luminosity or optical output with the GRB energy release for a set of 17 afterglows, 
5 of their afterglows falling in our plateau sample. For comparison, the slopes of the log-log 
fits obtained by Liang et al (2010): $L_p \propto E_\gamma^{1.17\pm 0.13}$ and $E_o \propto 
E_\gamma^{0.74\pm 0.10}$, are smaller than shown in Figure \ref{LpEg}.

 In the following subsections, we explore three scenarios for the generation of light-curve 
peaks and plateaus, and assess their ability to generate the steep $F_p \propto t_p^{-3}$ 
boundary exhibited by the peaky afterglow population.

\subsection{Passage of a spectral break}

 The passage of either the {\sl injection} frequency $\nu_i$, which is the peak frequency
of the $\nu F_\nu$ synchrotron spectrum, or of the {\sl cooling} frequency $\nu_c$ through 
the observing band causes a break in the afterglow light-curve. 

 {\sl Cooling frequency}.
 The afterglow spectrum has a shallow break at the characteristic synchrotron frequency $\nu_c$ 
of the electrons whose radiative cooling timescale equals the dynamical timescale,
across which the spectrum $F_\nu \propto \nu^{-\beta}$ softens by $\delta \beta = 1/2$
if $\nu_i < \nu_c$ and by $\delta \beta = 5/6$ if $\nu_c < \nu_i$. In the forward-shock model,
$\nu_c$ evolves slowly ($\nu_c \propto t^{-1/2}$ for a homogeneous medium and $\nu_c \propto 
t^{1/2}$ for a wind). Consequently, the passage of $\nu_c$ yields a weak steepening of the 
power-law flux $F_\nu \propto t^{-\alpha}$ by $\delta \alpha = \delta \beta |d\log \nu_c/d\log t|=
1/4$ or $5/12$ for $\nu_c > \nu_i$ and $\nu_c < \nu_i$, respectively, which is less than the 
break seen at the peaks and plateau ends of optical afterglows. 

 {\sl Injection frequency}.
 If $\nu_i < \nu_c$, the {\sl $\nu F_\nu$ peak frequency} $\nu_i$ is a stronger spectral break, 
from $F_\nu \propto \nu^{1/3}$ to $F_\nu \propto \nu^{-\beta}$ with $\beta \simeq 1$. 
For $\nu_c < \nu_i$, the afterglow spectrum has a weaker break at $\nu_i$, $F_\nu \propto 
\nu^{-1/2}$ to $F_\nu \propto \nu^{-\beta}$. The passage of $\nu_i$ through the optical
yields stronger light-curve break than $\nu_c$ because the evolution of $\nu_i$ is faster: 
$\nu_i \propto t^{-3/2}$. For $\nu_i < \nu_c$, the light-curve decay steepening at $\nu_i$ crossing
is $\delta \alpha = -(\beta+1/3) (d\log \nu_i/d\log t) = (3\beta+1)/2 \simeq 2$, and could account 
for an optical light-curve peak, provided that the medium is homogeneous. For $\nu_c < \nu_i$, 
$\delta \alpha = 1.5(\beta-1/2) \simeq 1.5$, and the $\nu_i$ crossing could produce only a plateau
end.  As a simple test of this scenario, the optical light-curve break produced by the passage 
of $\nu_i$ must be accompanied by a {\sl significant spectral softening of the afterglow optical 
spectrum}. 

 However, the passage of $\nu_i$ {\sl cannot be} the explanation for all optical light-curve
peaks or plateau ends shown in Fig \ref{z2} because
(1) peaky afterglows rise with an index $\alpha \in (-3,-2)$, i.e. they rise faster than what 
    can be accommodated by $\nu_i$ being above optical, and 
(2) variations in the afterglow parameters $E$ and $\epsilon_B$, which determine both the
    synchrotron peak flux $F_i$ at $\nu_i$ ($F_i \propto E \epsilon_B^{1/2}$ for a homogeneous 
    medium and $F_i \propto E^{1/2} \epsilon_B^{1/2}$ for a wind) and the epoch when 
    $\nu_i \propto E^{1/2} \epsilon_B^{1/2} t^{-3/2}$ crosses the optical, lead to positive 
    $F_p-t_p$ and $F_b-t_b$ correlations (in contrast, anticorrelations are observed).

\subsection{Misaligned jets}

 A misaligned jet yields an {\sl achromatic} light-curve peak when the jet has decelerated
enough for the observer's direction (at angle $\theta_o$ from the jet direction of motion) 
to enter the cone of opening $1/\Gamma$ into which the jet emission is relativistically beamed, 
$\Gamma$ being the jet Lorentz factor.

{\sl Possible failure}.
 This scenario can account for the fast rise of peaky optical afterglows, but has difficulty 
in explaining the slope of the $F_p-t_p$ anticorrelation shown in Figure \ref{z2}, if that 
relation arises from jets having various orientations $\theta_o$ relative to the observer.
In this framework, special relativity effects lead to a received flux 
$F_\nu \propto [2\Gamma/(\Gamma^2\theta_o^2+1)]^{\beta+3}$, hence, at the light-curve peak 
(when $\Gamma \theta_o = 1$), we have $F_p \propto \theta_o^{-(\beta+3)}$.
Rhoads (1999) has shown that the jet lateral spreading is dynamically significant when
$\Gamma \theta_o = 1$, it leads to an exponential jet deceleration $\Gamma \propto e^{-kr}$, 
and $t_p \propto r_o \theta_o^2$, where $r_o$ is the radius at which $\Gamma =\theta_o^{-1}$, 
after which the jet radius increases only logarithmically with observer time. 
Thus $t_p \propto \theta_o^2$, which together with $F_p \propto \theta_o^{-(\beta+3)}$ leads to
$F_p \propto t_p^{-0.5(\beta+3)}$.
Then, the observed anticorrelation slope of about $-3$ requires an optical spectral slope 
$\beta_o = 3$, which is much softer than measured typically ($\beta_o \simeq 1$), albeit at 
times after the optical peak.

{\sl Caveats}.
 However, the real $F_p-t_p$ anticorrelation could be less steep, perhaps only $F_p \propto
t_p^{-2}$, and consistent with the misaligned-jet model expectation for a $\beta_o = 1$
optical spectrum, for the following two reasons. First, there is an observational bias against
peaks occurring earlier than those shown in Figure \ref{z2}. Afterglows peaking at an earlier 
time must exist, as shown by those afterglows which display a decay from their first measurement. 
The other reason is that the sample of peaky afterglows shown in Figure \ref{z2} may be contaminated 
with early optical emission that does not arise from afterglow the blast-wave, but from the prompt 
emission mechanism (the GRB). Indeed, three of the brighter peaks shown in Figure \ref{z2} are 
coincident with a GRB pulse. In these cases, the true afterglow flux of the earlier light-curve 
peaks is lower than measured.
 
{\sl Contrived feature}.
 Misaligned jets can also account for plateau afterglows if the ejecta kinetic energy
per solid angle is not uniform within the jet opening. This is the model proposed by
Panaitescu \& Vestrand (2008) to explain the diversity of optical afterglow light-curves.
An unattractive aspect of this explanation is its contrived requirement of another jet 
producing the GRB emission, as the prompt emission from the misaligned optical jet is not 
relativistically beamed toward the observer. 

\subsection{Onset of deceleration}

\subsubsection{Peaky afterglows: impulsive ejecta release} 

 For an impulsive ejecta release and a narrow distribution of the ejecta Lorentz
factor after the burst phase, the forward-shock synchrotron emission light-curve prior to
the onset of deceleration displays a rise similar to that measured for peaky afterglows,  
and a decay rate compatible with the observations after the onset of deceleration.
Also, {\sl the onset of deceleration can yield a $F_p-t_p$ anticorrelation similar to that 
observed for peaky afterglows}, if that correlation arises from variations of the ambient 
density among afterglows, as shown below. 

\vspace*{2mm}
{\sl A. Homogeneous medium} \\
There are two cases where a decaying light-curve is obtained after the deceleration timescale 
$t_d \propto (E/n\Gamma_0^8)^{1/3}$ (which is the light-curve peak time): \\
(1) $F_{\nu_i<\nu_o<\nu_c} \propto E n^{(\beta+1)/2} \Gamma_0^{4\beta}$ and \\ 
(2) $F_{\nu_i,\nu_c<\nu_o} \propto E^{2/3} n^{(3\beta-1)/6} \Gamma_0^{4\beta-4/3}$. \\
In case (1), different ambient medium densities among afterglows yield
$F_p \propto t_p^{-1.5(\beta+1)}$, hence an optical spectral slope $\beta_o = 1$ is 
required to account for the observed anticorrelation, while the various initial Lorentz 
factors of afterglows lead to $F_p \propto t_p^{-1.5\beta}$, for which the measured 
anticorrelation requires a too soft optical spectrum, with slope $\beta_o = 2$. \\
In case (2), differences in $n$ or $\Gamma_0$ among afterglows induce the same 
anticorrelation: $F_p \propto t_p^{-(3\beta-1)/2}$, which requires a rather soft optical 
spectrum with slope $\beta_o = 7/3$ to account for the observed $F_p \propto t_p^{-3}$. \\
Thus, there is one scenario: \\
\centerline{$\nu_i<\nu_o<\nu_c$, $\beta_o=1$, various medium densities} \\ 
which can account for the measured $F_p-t_p$ anticorrelation displayed by peaky afterglows. 
Owing to the weak dependence of the deceleration timescale on the external density, $F_p
\propto n^{1/3}$ for $\beta_o = 1$, the 1.5 dex spread in the peak-time shown in Figure 
\ref{z2} requires that the ambient density varies by 4--5 dex.

\vspace*{2mm}
{\sl B. Wind-like medium} \\
 In this case, $t_d \propto E/(A\Gamma_0^4)$, where $A \propto (dM/dt)/v_w$ is the wind 
density parameter, which depends on the massive star (GRB progenitor) mass-loss rate 
$dM/dt$ and the wind terminal velocity $v_w$. The cases yielding a decaying light-curve 
after the onset of deceleration are: \\
(1) $F_{\nu_i<\nu_o<\nu_c} \propto E^{-\beta} A^{1.5\beta+1} \Gamma_0^{6\beta+2}$ and \\
(2) $F_{\nu_i,\nu_c<\nu_o} \propto E^{1-\beta} A^{(3\beta-1)/2} \Gamma_0^{6\beta-2}$. \\
 In case (1), diversity in $A$ leads to $F_p \propto t_p^{-(1.5\beta+1)}$, which requires
$\beta_o = 4/3$, while various $\Gamma_0$ among afterglows lead to $F_p \propto 
t_p^{-(3\beta+1)/2}$, which is the observed $F_p \propto t_p^{-3}$ for $\beta_o = 5/3$. \\
 In case (2), the result is the same as for a homogeneous medium: $F_p \propto 
t_p^{-(3\beta-1)/2}$ for either the $A$ or the $\Gamma_0$-induced correlation. \\
 Therefore, the deceleration-onset model accounts for the $F_p-t_p$ anticorrelation in the 
best way if \\
\centerline{$\nu_i<\nu_o<\nu_c$, $\beta_o = 4/3$, various wind densities} \\
hence $F_p \propto A^3$, and the wind density parameter $A$ should vary by 1.5 dex among afterglows.

\subsubsection{Unifying picture} 

 An extended release of ejecta, or one where the ejecta have a range of initial Lorentz 
factors after the burst, can lead to a gradual injection of energy into the blast-wave
because, in both scenarios, there will be some ejecta inner to the decelerating blast-wave,
which catch-up with the forward shock. In these scenarios, the forward-shock emission could 
be constant or slowly-decaying, followed by a steep decay when the energy injected stops being 
dynamically important. 
The correlation between the plateau flux and break epoch $F_b \propto t_b^{-1}$ shown in 
Figure \ref{z2} suggests a (very roughly) constant optical output among afterglows.

 Therefore, the unifying picture for the peaky and plateau optical afterglows shown in 
Figure \ref{z2} is that they all arise from the onset of the blast-wave deceleration, 
the former being identified with an impulsive ejecta release where all ejecta have
the same Lorentz factor after the burst phase, and the latter being attributed to
energy injection in the forward-shock due to $(i)$ an extended ejecta release, 
$ii)$ a wide distribution of the ejecta initial Lorentz factor, or both.

 In the next section, we compare the optical and X-ray light-curves of the identified
two classes of optical afterglows, to search for differences in the correlation of the
optical and X-ray flux histories, and to identify the models that can account for the diversity
of optical vs. X-ray light-curve behaviours.

\section{Optical and X-ray afterglows}

\subsection{Expectations}

 In the forward-shock model, the flux power-law decay indices at optical and X-ray frequencies 
are equal if there is no spectral break between these two domains. If $\nu_o < \nu_c < \nu_x$, 
then the decay indices difference is $\alpha_x - \alpha_o = 1/4$ (the X-ray flux decays 
faster the optical flux) for a homogeneous medium, and $\alpha_x - \alpha_o = -1/4$ 
(the optical flux falls-off faster than the X-ray's) for a wind-like medium. 

 Three caveats to the above results are worth mentioning. 

 First is that these results stand when the electron radiative cooling is synchrotron-dominated.
If it is due mainly 
to inverse-Compton losses, then the fastest possible evolutions of the cooling break are 
$\nu_c \propto t^{1/2}$ (thus $\alpha_x - \alpha_o = -1/4$) for a homogeneous medium and 
$\nu_c \propto t^{5/2}$ (hence $\alpha_x - \alpha_o = -5/4$), but the required conditions 
for this case to occur are unlikely to be maintained over a long time. 

\begin{figure*}
\psfig{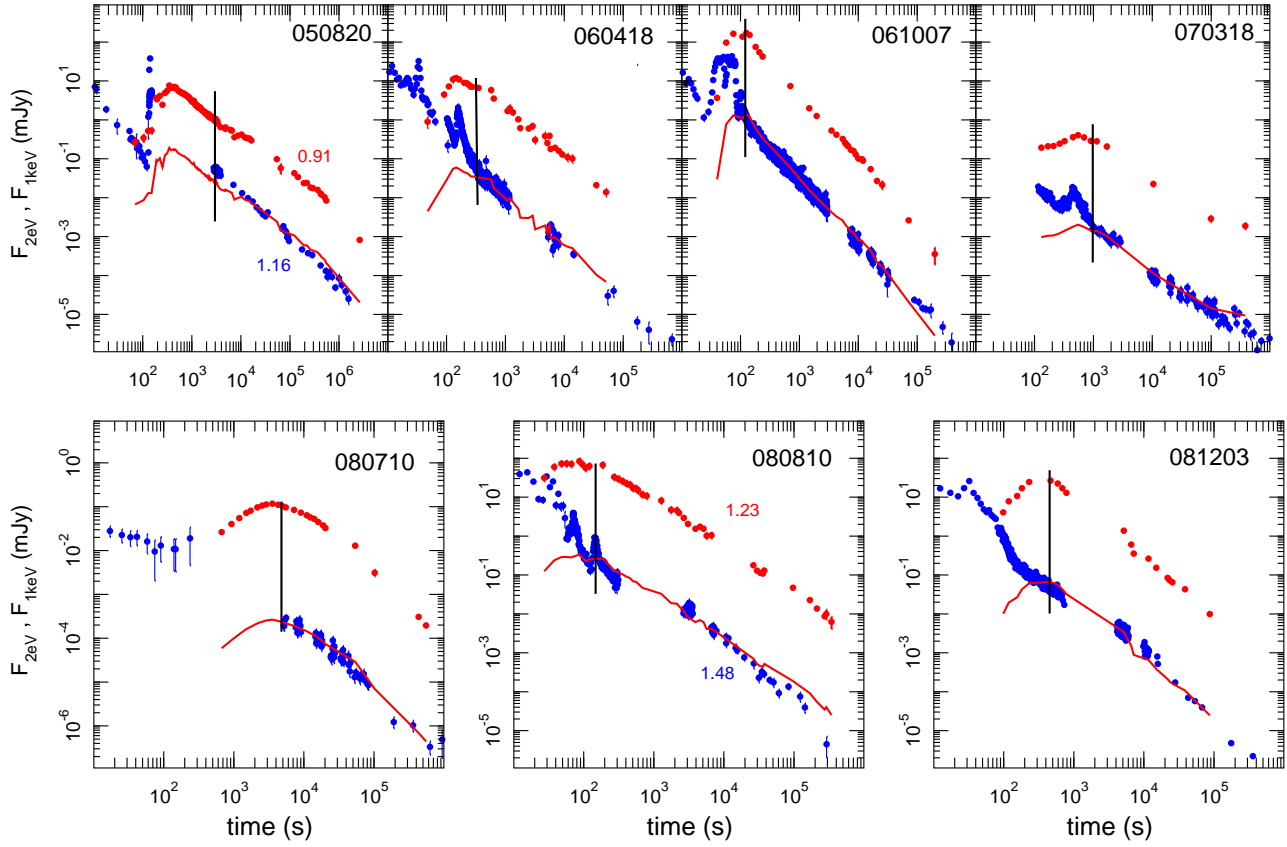}
\caption{ Optical (red symbols) and X-ray (blue symbols) light-curves of afterglows
  with optical {\bf peaks} and {\bf coupled} optical and X-ray light-curves.
  To illustrate that the two frequencies light-curves follow each other, we also
  show the optical light-curve shifted to overlap the X-ray light-curve. The X-ray
  light-curves do not exhibit a peak at the same time with the optical because
  the optical peaks occur early, when the X-ray emission is still dominated by the 
  decaying contribution from the prompt (burst) mechanism. Vertical lines indicate the 
  first X-ray measurement, after which the prompt mechanism contribution is negligible 
  and the X-ray flux is from the afterglow. }
\label{peak1}
\end{figure*}

 The second caveat is that a slowly decreasing or constant optical flux results also for 
$\nu_o < \nu_i$. In this case, an optical light-curve plateau is obtained if:\\
(1) $\nu_c < \nu_o < \nu_i$ and the medium is homogeneous, for which $\alpha_o = 1/4$ and the 
  optical spectral slope is $\beta_o = 1/2$, \\
(2) $\nu_o < \nu_c < \nu_i$ and the medium is a wind, for which $\alpha_o = 2/3$ and $\beta_o = -1/3$, \\
(3) $\nu_o < \nu_i < \nu_c$ and the medium is a wind, for which $\alpha_o = 0$ and $\beta_o = -1/3$.
We note that a rising afterglow optical spectrum has never been observed (to the best of 
our knowledge), hence only case (1) above may be relevant.

 The third caveat is that, if energy is added to the forward-shock, then the evolution of 
the cooling frequency is "accelerated" and the difference between the optical and X-ray 
flux decay indices is larger. If the increase of the blast-wave kinetic energy is described 
by $E \propto t^e$, then $\nu_c \propto t^{-(e+1)/2}$ and $\alpha_x - \alpha_o = (e+1)/4$ 
for a homogeneous medium. For a wind medium, $\nu_c \propto t^{(e+1)/2}$ and $\alpha_x - 
\alpha_o = -(e+1)/4$.

 Using the above expectations for the forward-shock model, we define {\sl coupled} optical 
and X-ray light-curves those for which $|\alpha_x - \alpha_o|$ is {\sl zero, 1/4, or slightly 
larger than 1/4}, and define {\sl decoupled} light-curves those for which $|\alpha_x - \alpha_o|$ 
is {\sl substantially larger than 1/4}, with the caveat that a decay index difference $\alpha_x - 
\alpha_o = (3p-3)/4 = (6\beta_x-3)/4 \simeq 3/4$ could be obtained for the forward-shock light-curves
in the $\nu_c < \nu_o < \nu_i < \nu_x$ case, if the circumburst medium is homogeneous. 
Additionally, {\sl decoupled} light-curves are also those which display a chromatic light-curve 
breaks, present at only one frequency.

\begin{figure*}
\psfig{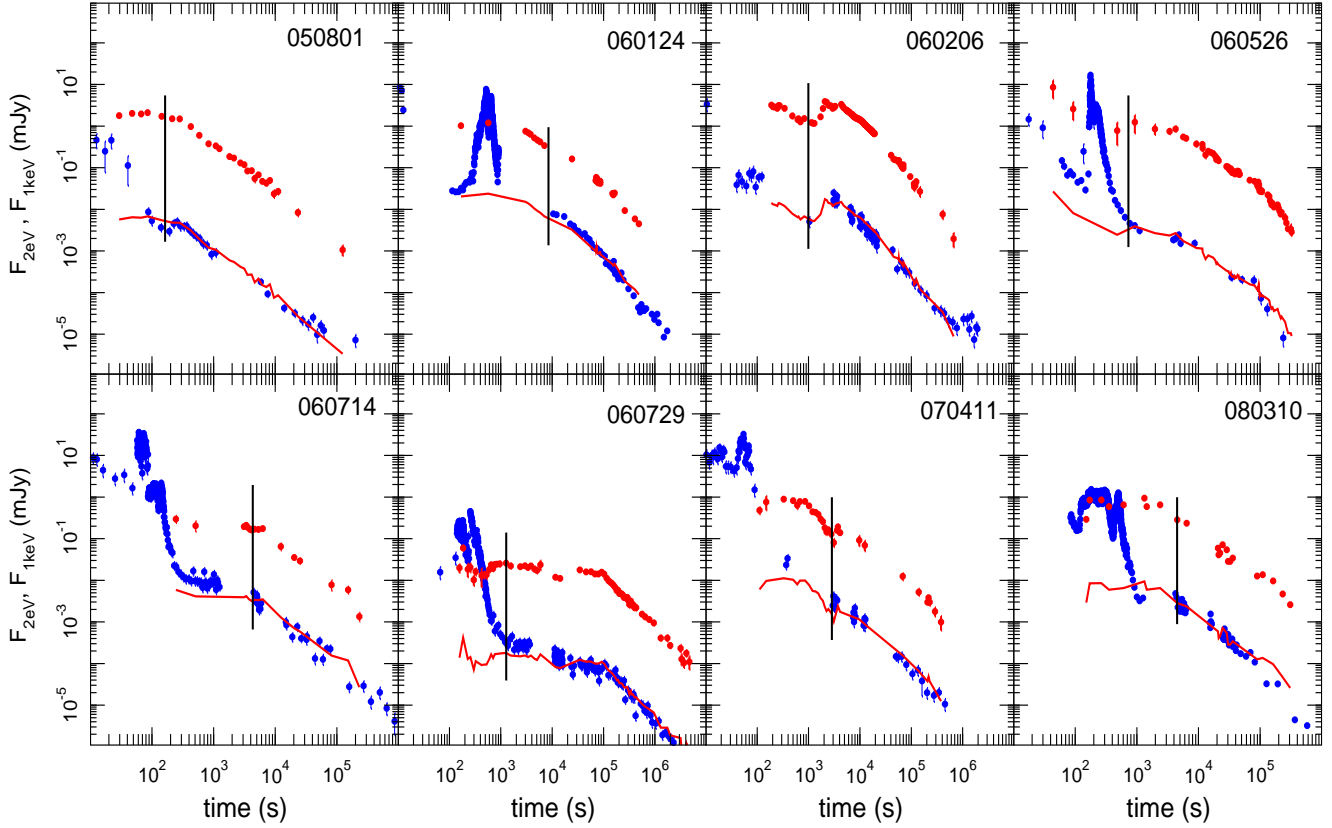}
\caption{ Afterglows with optical {\bf plateaus} and {\bf coupled} optical and X-ray 
   light-curves. Other details as for Figure \ref{peak1}. }
\label{plat1}
\end{figure*}

 Figures \ref{peak1}--\ref{plat2} show the optical and X-ray light-curves for all afterglows 
with optical peaks/plateaus 
and with a good temporal coverage in the X-ray, which we were able to collect as of July 2010.
The implications of their coupling is discussed below.

\subsection{Afterglows with coupled light-curves}

 The majority of coupled light-curves shown in Figures \ref{peak1} and \ref{plat1} have 
$\alpha_o = \alpha_x$, as can be seen from the good overlap of the shifted optical light-curves 
and the X-ray light-curve. In these cases, the cooling frequency of the synchrotron spectrum 
is not between optical and X-ray. Different optical and X-ray flux decay indices are found for 
the peaky afterglows 050820 and 080810 (Fig \ref{peak1}), for which $\alpha_x -\alpha_o = 0.25 
\pm 0.03$, indicating that $\nu_o < \nu_c < \nu_x$, that the ambient medium is homogeneous, 
and that no significant energy injection took place (as expected for peaky afterglows).

 The well-coupled light-curves of the 14 afterglows shown in Figures \ref{peak1} and \ref{plat1} 
indicate that a single dissipation mechanism (forward-shock) and radiating process (synchrotron) 
produce the afterglow emission at both frequencies. 
 In contrast, Figures \ref{peak2} and \ref{plat2} show afterglows whose light-curves display 
substantially different decays in the optical and X-ray or have chromatic light-curve breaks. 
The possible reasons for such decoupled light-curves are discussed below.

\subsection{Decoupled light-curves without chromatic breaks}

 Decoupled light-curves without chromatic breaks, such as those of GRB afterglows 080129, 
090423, and 090510 (Fig \ref{plat2}), may not require a 
different origin of the optical and X-ray emissions. Instead, it may be that optical is 
synchrotron and X-rays are from inverse-Compton scattering in the forward-shock.
 For an energy injection law $E \propto t^e$, a power-law distribution of electrons with 
energy $dN/d\gamma \propto \gamma^{-p}$, and a homogeneous medium, the power-law index
of the synchrotron (optical) power-law decaying flux is
\begin{equation} 
 \alpha_o = \frac{1}{4} \left\{  \begin{array}{ll}
       3p-3-(p+3)e & \nu_i^{(sy)} < \nu_o < \nu_c^{(sy)} \\
       3p-2-(p+2)e & \nu_c^{(sy)} < \nu_o  
 \end{array} \right. \;,
\end{equation}
while for the inverse-Compton (X-ray)
\begin{equation} 
 \alpha_x = \frac{1}{8}\left\{  \begin{array}{ll}
       9p-11-(3p+7)e & \nu_i^{(ic)} < \nu_x < \nu_c^{(ic)} \\
       9p-10-(3p+2)e & \nu_c^{(ic)} < \nu_x 
 \end{array} \right. \;.
\end{equation}
For a wind medium
\begin{equation} 
 \alpha_o = \frac{1}{4} \left\{  \begin{array}{ll}
       3p-1-(p+1)e & \nu_i^{(sy)} < \nu_o < \nu_c^{(sy)} \\
       3p-2-(p+2)e & \nu_c^{(sy)} < \nu_o 
 \end{array} \right. \;,
\end{equation}
\begin{equation} 
 \alpha_x = \left\{  \begin{array}{ll}
       p- (p-1)e/2 & \nu_i^{(ic)} < \nu_x < \nu_c^{(ic)} \\
       p-1- pe/2 & \nu_c^{(ic)} < \nu_x   
 \end{array} \right. \;.
\end{equation}
When there is no energy injection, the previous equations give the flux decay index by setting $e=0$. 

 With the above results, it can be shown that the synchrotron self-Compton (SsC) model can explain 
the decoupled light-curves of the following afterglows: \\
$i)$ 080129 -- $\nu_o < \nu_c^{(sy)}$, $\nu_x < \nu_c^{(ic)}$, $e \simeq 2.0$ (for an assumed 
    $p=2.5$), \\
$ii)$ 090423 -- $\nu_o < \nu_c^{(sy)}$, $\nu_x < \nu_c^{(ic)}$, $p \simeq 1.6$, $e \simeq 1.0$, and \\
$iii)$ 090510 --  $\nu_o < \nu_c^{(sy)}$, $\nu_x < \nu_c^{(ic)}$, $p \simeq 2.2$, $e \simeq 2.4$
 (but here the resulting rise and decay are slightly faster than measured), \\
if energy injection is negligible after the achromatic light-curve break and if the ambient medium 
is a wind.

\begin{figure*}
\psfig{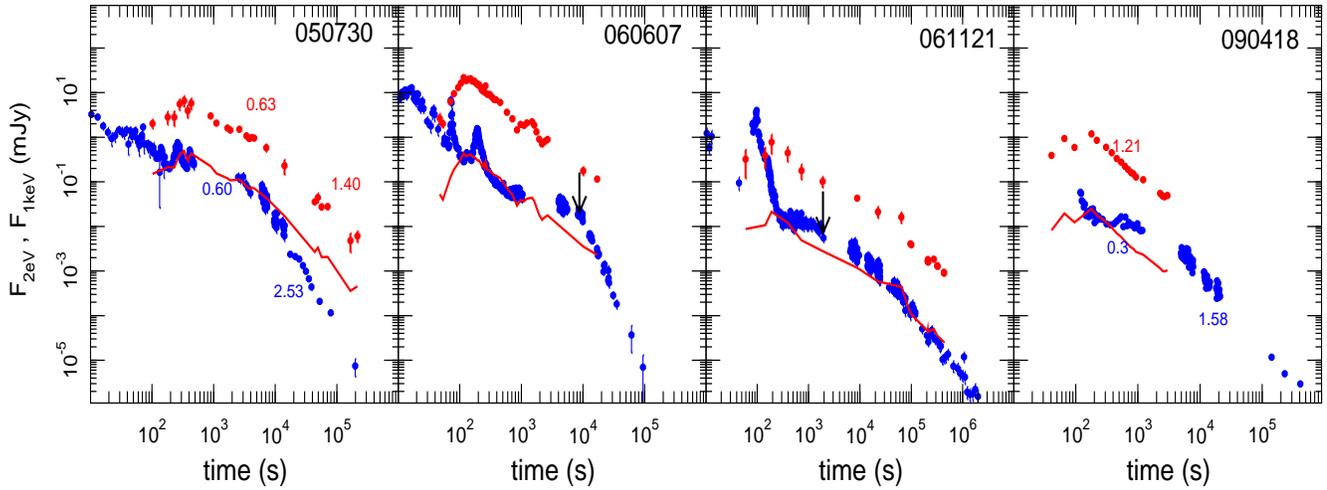}
\caption{ Afterglows with optical {\bf peaks} and {\bf decoupled} optical and X-ray 
  light-curves. Also shown is the optical light-curve shifted to match the X-ray flux at 
  the earliest time when the X-ray flux could be predominantly from the afterglow. 
  For 050730, the decoupling of light-curves occurs after 10 ks, when the indices of the 
  optical and X-ray flux power-law decays differ by more than the of 1/4 allowed by the 
  standard synchrotron forward-shock model. Afterglows 060607 and 061121 display a 
  chromatic X-ray light-curve break at the time indicated with an arrow. The light-curves
  of afterglow 090418 display different decays at 0.3--1 ks. }
\label{peak2}
\end{figure*}

 For the peaky afterglows 050730 and 090418 (Fig \ref{peak2}), we cannot find a set of parameter 
$(p,e)$ to accommodate their decoupled optical and X-ray decays within the SsC model with energy 
injection cessation at the light-curve break epoch (model is overconstrained, with 2 free parameters
and 3-4 observational constraints), which suggests that either energy injection continues after 
the break (but with a different power-law exponent $e$), or a different origin of the decoupled 
optical and X-ray light-curves.

\subsection{Decoupled light-curves with chromatic breaks}

\begin{figure*}
\psfig{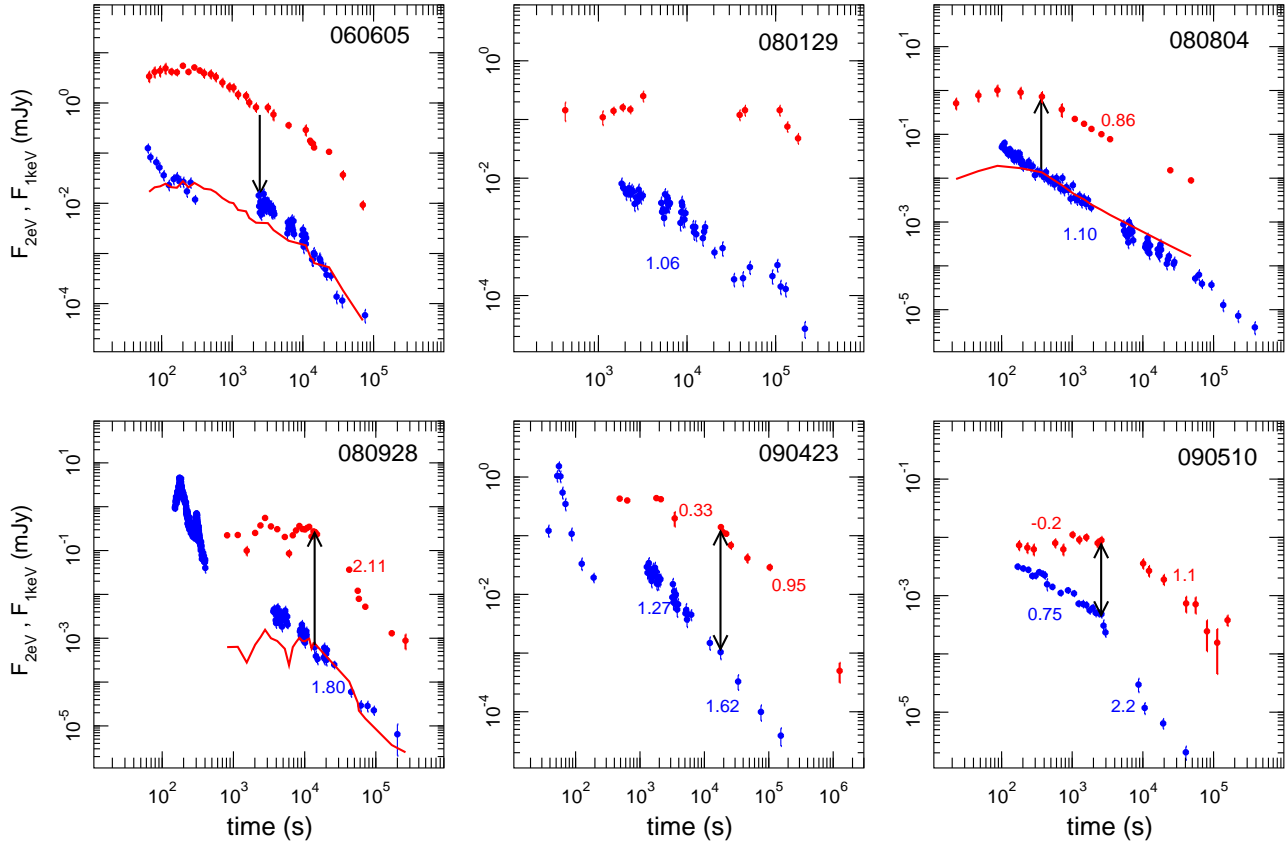}
\caption{ Afterglows with optical {\bf plateaus} and {\bf decoupled} optical and X-ray
   light-curves. 060605 must have had a chromatic X-ray light-curve break around at a few ks.
   080804 and 080928 display chromatic optical light-curve breaks, which may be due to the
   synchrotron spectral peak crossing the optical, consistent with the post-plateau decay 
   indices being the same. 080129, 090423, and 090510 show very different decays, with the 
   X-ray light-curve decaying much faster than the optical, as expected for the forward shock
   emission and a wind-like medium if the X-ray are inverse-Compton and optical is synchrotron.
   The achromatic light-curve breaks of 090423 and 090510 indicate an origin in the
   blast-wave dynamics, and could arise from cessation of energy injection into the 
   blast-wave. Numbers indicate the flux power-law decay index.  }
\label{plat2}
\end{figure*}

 {\sl Chromatic optical light-curve breaks}.
 They are observed for the plateau afterglows 080804 and 080928 (Fig \ref{plat2}) and may originate 
from the peak $\nu_i$ of the synchrotron spectrum crossing the optical. In this scenario, the 
optical and X-ray light-curves should be coupled after the break. A simple test for the origin 
of the optical light-curve break in the passage of $\nu_i$ is the softening of the optical 
spectrum across the break epoch. 

 {\sl Chromatic X-ray light-curve breaks}.
Such breaks are observed for the peaky afterglows 060607 and 061121 (Fig \ref{peak2}), and for 
the plateau afterglow 060605 (Fig \ref{plat2}). They suggest that the optical and X-ray emissions 
either ($i)$ arise from different radiation processes or $ii)$ have a different origin. 
The former scenario is an unlikely explanation because, in this case, the X-ray light-curve 
break can be produced only by a spectral break crossing the observing band, which is in 
contradiction with the general lack of an X-ray spectral evolution observed to occur simultaneously 
with a chromatic light-curve break (e.g. Nousek et al 2006). The latter scenario is motivated 
by that, if the X-ray and optical afterglow emissions had the same origin, then a light-curve 
break arising from a change in the blast-wave dynamics should also be present in the optical 
light-curve. 

 Reprocessing of the lower-frequency forward-shock emission through up-scattering by another part 
of the outflow can yield X-ray light-curves that are decoupled from the optical provided that 
$i)$ the scattering outflow has an optical thickness (well below unity) sufficiently large, to account 
for the measured X-ray flux, and 
$ii)$ the scattering outflow moves at a higher Lorentz factor than the forward shock, to boost 
the shock's emission to higher frequency and to yield a flux larger than that coming directly 
from the forward shock (Panaitescu 2008).

 A simple expectation for the bulk-scattering model is that chromatic X-ray light-curve breaks 
should be seen more often for afterglows with optical plateaus than for peaky afterglows. That 
is so if, as proposed here, afterglows with plateaus arise from an incoming outflow that adds 
energy to the blast-wave, as that outflow upstairs the forward-shock emission. In contrast,
if afterglows with peaks arise from impulsive ejecta releases, then an outflow inner to the 
forward-shock should not exist for them. 

 Contrary to that expectation, only one plateau afterglow (060605) displays a chromatic 
X-ray light-curve break, while two peaky afterglows (060607 and 061121) have such a feature. 
Taken at face value, the deceleration-onset model for peaky and plateau afterglows is in 
contradiction with this low-number "statistics" and, evidently, a larger number of optical 
and X-ray afterglows should be studied before drawing a reliable conclusion.
Nevertheless, we note that an outflow inner to the blast-wave may also exist for peaky afterglows
and, while that outflow could carry too little energy to modify the forward-shock dynamics and 
yield an optical light-curve plateau instead of a peak, the upscattered emission from the incoming 
outflow could still overshine the direct forward-shock emission and produce a chromatic X-ray 
light-curve plateau.

\subsection{X-ray to optical flux ratio}

\begin{figure*}
\centerline{\psfig{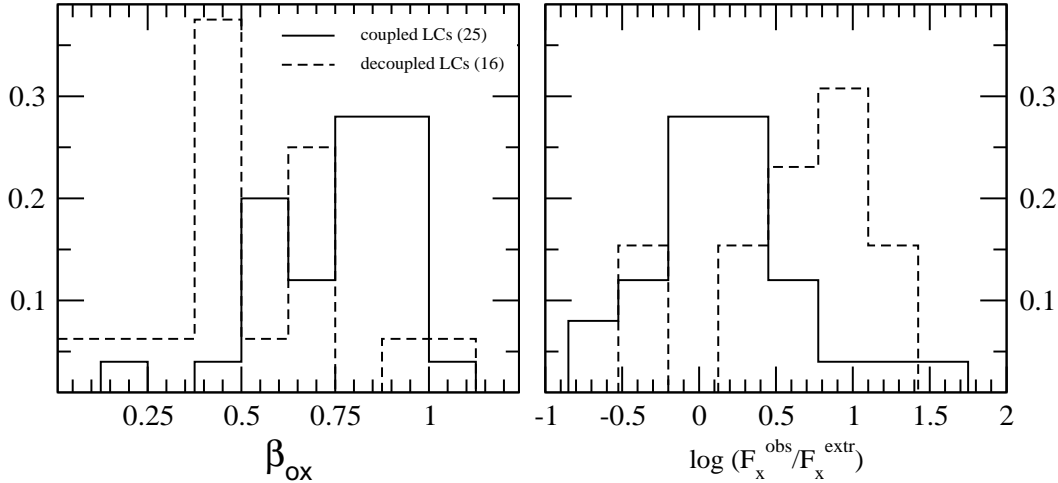}}
\caption{ Distributions of the optical-to-X-ray slope $\beta_{ox} = - \log (F_x/F_o)/
  \log (\nu_x/\nu_o)$ for 25 afterglows with coupled optical and X-ray light-curves (solid line) 
  and 16 decoupled afterglows (dashed line). 25 of these 41 afterglows are shown in Figures 
  \ref{peak1}--\ref{plat2}; to them, we have added another 16 whose optical light-curves displays 
  a decay from the earliest measurement (i.e. the peak or plateau phase were missed). 
  For afterglows with a single power-law light-curve, the fluxes $F_x$ and $F_o$ used here are 
  around 10 ks, while for afterglows with  broken power-law light-curves, the fluxes are at the 
  epoch of the break. The average spectral slope of decoupled and coupled afterglows is 
  $\bar{\beta}_{ox} = 0.52 \pm 0.25$ and $\bar{\beta}_{ox} = 0.76 \pm 0.20$, respectively. 
  $\chi^2$ statistics gives a 2 percent probability that the two distributions are drawn from 
  the same parent distribution.  }
\label{betaox}
\end{figure*}

 Figure \ref{betaox} shows the distributions of the optical-to-X-ray spectral slope $\beta_{ox} = -
\log (F_x/F_o)/\log (\nu_x/\nu_o)$ for afterglows with coupled (same decays, without breaks or 
with achromatic breaks) and decoupled (different decays or with chromatic breaks) optical and
X-ray light-curves. 
The simple result shown in Figure \ref{betaox} is that the afterglows with decoupled light-curves
are, on average, brighter in the X-ray relative to the optical (smaller slope $\beta_{ox}$)
than the afterglows with coupled light-curves. Quantitatively, the difference between the average 
slopes $\bar{\beta}_{ox}$ of the two categories of afterglows is $\delta \bar{\beta}_{ox} = 0.24 
\pm 0.32$, which means that the X-ray-to-optical flux ratio of decoupled afterglows is, on average, 
$(\nu_x/\nu_o)^{\delta \bar{\beta}_{ox}} = (\rm{1\,keV/2\,eV})^{0.24} = 4.4$ times larger than 
for the afterglows with coupled light-curves.

 A similar conclusion is reached if, instead of the optical flux, we normalize the measured X-ray 
flux to that estimated from the optical flux assuming that the afterglow spectrum were a pure 
power-law. Here, we define $y \equiv \log (F_x^{(obs)} /F_x^{(extr)})$ with $F_x^{(extr)} = 
F_o^{(obs)} (\nu_o/\nu_x)^{\beta_o}$, where $\beta_o$ is the intrinsic power-law slope of the 
optical spectrum. As $\beta_o$ is hard to determine 
accurately (owing primarily to the unknown reddening by dust in the host galaxy), we use the 
value required within the forward-shock model by the measured power-law optical flux decay 
$F_o^{(obs)} \propto t^{-\alpha_o}$, with $\alpha_o$ measured after the optical plateau 
(i.e. when the blast-wave is adiabatic): $\beta_o = (2/3)\alpha_o + c$. 
 The constant $c$ depends on the location of optical relative to the cooling frequency $\nu_c$ 
of the forward-shock synchrotron emission spectrum and, for $\nu_o < \nu_c$, also on 
the radial stratification of the ambient medium. The major assumption made here is that the
constant $c$ is the same for all afterglows.
For instance, if the ambient medium is homogeneous and if $\nu_o < \nu_c$ (for which $c=0$), 
we find that $\bar{y} = 0.22 \pm 0.54$ for afterglows with coupled light-curves and $\bar{y} = 
0.64 \pm 0.51$ for afterglows with decoupled light-curves, the difference between the average 
values being $\delta \bar{y} = 0.42 \pm 0.74$. This indicates that the X-ray flux of decoupled 
afterglows is, on average, $10^{\delta \bar{y}} = 2.6$ times brighter than for coupled afterglows. 
We note that $\bar{y}$ depends on the constant $c$, but $\delta \bar{y}$ does not, hence this 
conclusion is independent of the assumed optical spectral regime and medium stratification.

 That the afterglows with decoupled light-curves have a higher X-ray-to-optical flux ratio 
indicates that there is a mechanism which operates mostly/only in decoupled afterglows and which 
yields more X-ray emission than the underlying synchrotron flux from the forward shock. That 
mechanism could be local inverse-Compton scattering or external upscattering (bulk and/or 
inverse-Compton). Alternatively, the higher X-ray brightness of decoupled afterglows could
be explained if there were a process that reduces the optical emission of coupled afterglows. 
Given that the media involved in GRB afterglows are optically thin to electron scattering,
that reduction of the optical flux should be due to dust extinction in the host galaxy.
However, there is no obvious reason for which afterglows with coupled light-curves should
undergo more dust extinction than decoupled afterglows.

 We note that the optical luminosity distributions of afterglows with coupled and decoupled 
light-curves are consistent with being drawn from the same parent distribution, a conclusion
which also holds for the X-ray luminosities of these two classes.
For all afterglows moved at same redshift $z=2$, we find that the average optical flux (in mJy) 
at a fixed time (10 ks) is $<\log F_o>_c = -1.03 \pm 0.61$ for coupled afterglows and $<\log F_o>_d
= -1.35 \pm 0.62$ for decoupled afterglows. For the $z=2$ X-ray flux (in $\mu$Jy), $<\log F_x>_c = 
0.02 \pm 0.60$ and $<\log F_x>_d = 0.15 \pm 0.77$. Thus, decoupled afterglows are on average only 
slightly dimmer (by factor 2.0) in the optical than coupled afterglows and only slightly brighter 
(by a factor 1.3) in the X-ray, the difference between the two types of afterglows becoming clearer 
in the distributions of the X-ray-to-optical flux ratio. 


\section{Conclusions}

 Based on their light-curves appearance, we identify two types of optical afterglows, with 
light-curve peaks and with plateaus (Figure \ref{z2}). 
 In the plateau end luminosity--epoch plane, plateau afterglows show a $F_b \propto t_b^{-1}$ 
anticorrelation that might suggest an universal afterglow optical output. However, the optical
energy release during the plateau, shown in Figure \ref{LpEg}, has a substantial spread.  

 In the peak luminosity--peak epoch plane, the distribution of peaky afterglows displays a 
$F_p \propto t_p^{-3}$ bright edge, with earlier and/or dimmer peaks being missed. We speculate
that this steep dependence seen at the bright edge arises from the mechanism that generates 
optical light-curve peaks and we have discussed/investigated three such possible mechanisms.

 One is that optical observations are made below the synchrotron spectrum peak $\nu_i$, which
can also yield light-curve plateaus. This mechanism is easily invalidated by that $\nu_i$ crossing 
the optical should lead to a positive correlation of the peak/plateau end flux with epoch, 
contrary to the observed anticorrelations. 

  In our previous study (Panaitescu \& Vestrand 2008), we have attributed peaky afterglows 
to the observer being slightly outside the jet initial aperture and plateau afterglows to 
outflows endowed with an angular structure and an off-axis observer location. 
For a typical optical afterglow spectrum ($F_\nu \propto \nu^{-1}$), variations in the observer 
offset angle induce a $F_p-t_p$ anticorrelation that is less steep than measured. 
 However, there are two observational selection effects that could make the slope of the observed 
anticorrelation appear steeper. 
 First is that rapidly slewing telescopes capable of observing during the first few minutes 
after a GRB are typically less sensitive than those observing the afterglow at later stages. 
That means that faint, early peaks located in the left-lower corner of Figure \ref{z2} are missed.
 The second factor that could make the sample of early peaks appear brighter is contamination 
from the prompt mechanism. 
 A more troubling issue with this model is the required and contrived double-jet structure of 
the ejecta which it requires, with one jet moving toward the observer and producing the burst 
emission, and another slightly offset jet producing the afterglow emission and becoming visible 
after it has decelerated enough.

 Here, we propose another model to unify peaky and plateau afterglows, in which the two classes 
are related to the duration of the central engine, the light-curve peak and plateau break corresponding
to the end of energy injection in the blast-wave and to the onset of its deceleration. 
 In this model, plateau afterglows are associated with an extended production of ejecta or with an 
impulsive one but with a range of ejecta initial Lorentz factors, such that energy injection in the 
forward-shock can last up to 100 ks, while peaky afterglows are associated with an impulsive release 
of ejecta that have a narrow distribution of Lorentz factor.\footnotemark 

 We find that the exponent of the $F_p-t_p$ correlation is set by the slope of the afterglow 
optical spectrum ($F_o \propto \nu^{-\beta_o}$). That slope is very rarely measured at the early times 
when the optical peaks occur; at later times, $\beta_o \in (0.7,1)$ for most of the peaky afterglows 
shown in Figure \ref{z2}. The hardest optical spectrum for which variations in the afterglow parameters 
that set the deceleration/peak timescale yield the observed $F_p \propto t_p^{-3}$ correlation is 
$\beta_o = 1$, obtained for a homogeneous ambient density. 
Then, the observed range of optical peak fluxes (or peak epochs) requires the external density to vary 
by 4-5 dex among the afterglows shown in Figure \ref{z2}. 
\footnotetext{ If the burst emission arises from internal shocks, then the narrow distribution of ejecta 
Lorentz factors should occur only after the burst phase, as otherwise the internal shocks will have 
a very low dissipation efficiency and the GRBs of peaky afterglows should be much dimmer than those 
of afterglows with plateaus. This is in contradiction with the GRB energies shown in Figure \ref{LpEg}. }
 
 This model for the dichotomy of optical light-curves has a straightforward implication.
 For peaky afterglows arising from impulsive ejecta releases, it is quite likely that all the 
afterglow ejecta participated in the production of the burst emission.
 As some optical plateaus may arise from a long-lived central engine, it is possible that most of 
the afterglow ejecta did not produce the short-lived GRB emission.
 Thus, the {\sl optical output of afterglows with plateaus is expected to be less correlated with 
the burst energy than for peaky afterglows}. That seems to be, indeed, the case (Figure \ref{LpEg}). 

 We have compared the optical and X-ray light-curves of peaky and plateau afterglows and have 
found that, for 15 of the 25 afterglows with good coverage, the light-curves at these two 
frequencies are well-coupled, having the same power-law decay indices and achromatic breaks. 
For the other 10, we propose the following scenarios to explain their decoupled light-curves: \\
(1) if the external medium is sufficiently dense, then local (inverse-Compton) scattering may
overshine in the X-rays the synchrotron emission from the forward shock. This synchrotron
self-Compton model can account for the decoupled X-ray and optical flux decays of GRB afterglows 
080129, 090423, and 090510, but has difficulties in accounting for the light-curves of
afterglows 050730 and 090418. In this scenario, the X-ray (inverse-Compton) afterglow flux is
expected to decrease faster than the optical (synchrotron) flux. \\
(2) if the synchrotron spectrum peak passes through the observing band, then a chromatic light-curve 
break will result, which is more likely to be observed at optical frequencies, as for afterglows 
080804 and 080928. In this scenario, the chromatic light-curve break should be accompanied by a 
strong spectral softening. \\
(3) if a sufficiently relativistic and pair-rich outflow exists behind the forward-shock,
then bulk-scattering of the forward-shock emission by this incoming outflow may overshine the
direct emission from the forward shock and yield a chromatic X-ray light-curve break, as observed 
for GRB afterglows 060605, 060607, and 061121. In this model, the bulk-scattered emission mirrors
the radial distribution of the Lorentz factor and/or optical thickness of the scattering outflow, 
hence the light-curves of the scattered (X-ray) and direct (optical) emissions are naturally decoupled. 

\end{document}